\documentclass[CEJP,DVI]{cej} % use DVI command to enable LaTeX driver
\usepackage{layout}
\usepackage{amsmath}
\usepackage{textcomp}
\usepackage{hyperref}
\usepackage{graphicx}
\usepackage{sidecap}
\title{Highlights of the Beam Energy Scan from STAR}

\articletype{Review Article}

\author{A. Schmah for the STAR Collaboration\inst{}\email{aschmah@lbl.gov}}

\institute{
     \inst{} Lawrence Berkeley National Laboratory,\\
     1 Cyclotron Road, Berkeley, CA 94720, USA
          }

\abstract{The first part of the beam energy scan (BES) program at RHIC was successfully completed in the years 2010 and 2011. First STAR results from particle yield measurements are in good agreement with previously published data from SPS  and AGS experiments whereas other results like azimuthal HBT and $K/\pi$ event-by-event fluctuations differ at some energies. In addition, new observations like the centrality dependence of chemical freeze-out parameters ($T_{\rm{ch}}$ and $\mu_{B}$) or the smoothly increasing difference with decreasing energy in the elliptic flow $v_{2}$ between particles and corresponding anti-particles, are discussed. 
}


\begin{document}
\maketitle

%% ###################################################################

\section{Introduction}

Several heavy-ion measurements from the past decade indicate the existence of a QCD phase with quasi free quarks and gluons at high energy A+A collisions \cite{STAR_ncq}. The quark-gluon plasma (QGP) should have fundamentally different characteristics compared to the hadron gas phase which appears at lower energies. One of the goals of the beam energy scan (BES) at the Relativistic Heavy-Ion Collider (RHIC) facility is to find direct signatures for a transition between these two phases \cite{STAR_BES1,STAR_BES2}. For this purpose, a series of Au+Au experiments at different energies were accomplished in the years 2010 and 2011 at  RHIC.  Table \ref{table_1} gives an overview of the various beam energies and the collected statistics of minimum bias events at STAR.
Within the range of the chosen BES energies it is expected that fireballs are created which properties are around the transition line. For each of these energies multiple observables like azimuthal anisotropies ($v_{n}$), $p_{T}$ spectra, particle yields and ratios, azimuthal HBT,  $R_{\rm {CP}}$ etc. are studied. A rapid change at a certain energy of one or more of these observables could indicate an underlying phase transition. In addition, the disappearance of QGP signatures like number-of-constituent-quark (NCQ) scaling are studied as a function of energy. Comparisons to model predictions are performed as well, to identify an agreement or significant differences with the data.
Lattice QCD calculations show a crossover between the hadron gas and the QGP phase at low baryon chemical potentials $\mu_{B}$ \cite{ref_lattice}, whereas a phase transition is expected at higher $\mu_{B}$ values. This would automatically result in a QCD critical point. To find signatures of this critical point is another important goal of the BES program. For this purpose event-by-event fluctuation measurements are performed such as the net proton or $K/\pi$ measurements. 

\begin{table}[htc]
\centering
\footnotesize\rm
\begin{tabular}{ c  c   }
Energy (GeV)&Used MB-statistics ($10^{6}$)\\
\hline
\hline
7.7&4.3\\
11.5&11.7\\
19.6&35.8\\
27&70.4\\
39&130.4\\
62.4&67.3\\
\end{tabular}
\caption{\label{table_1}Number of collected minimum bias (MB) Au+Au events at STAR.}
\end{table}

\section{Results}

In the following sections, a selection of BES results from the STAR experiment is highlighted. The first part shows the energy dependence of the chemical freeze-out conditions and $\phi$-meson  $R_{\rm {CP}}$. In the next section elliptic flow and azimuthal HBT results are presented. The last section shows results from event-by-event $K$/$\pi$ and $p$/$\pi$ dynamical fluctuations.    

\subsection{Chemical freeze-out conditions and $\phi$-meson  $R_{\rm {CP}}$}

\begin{figure}
  %\subfloat{\label{fig_Tchem}\includegraphics[width=0.335\textwidth]{Tchem_cent.eps}} \qquad               
  %\subfloat{\label{fig_PhiRCP}\includegraphics[width=0.37\textwidth]{Phi_RCP.eps}}
\includegraphics[width=0.37\textwidth]{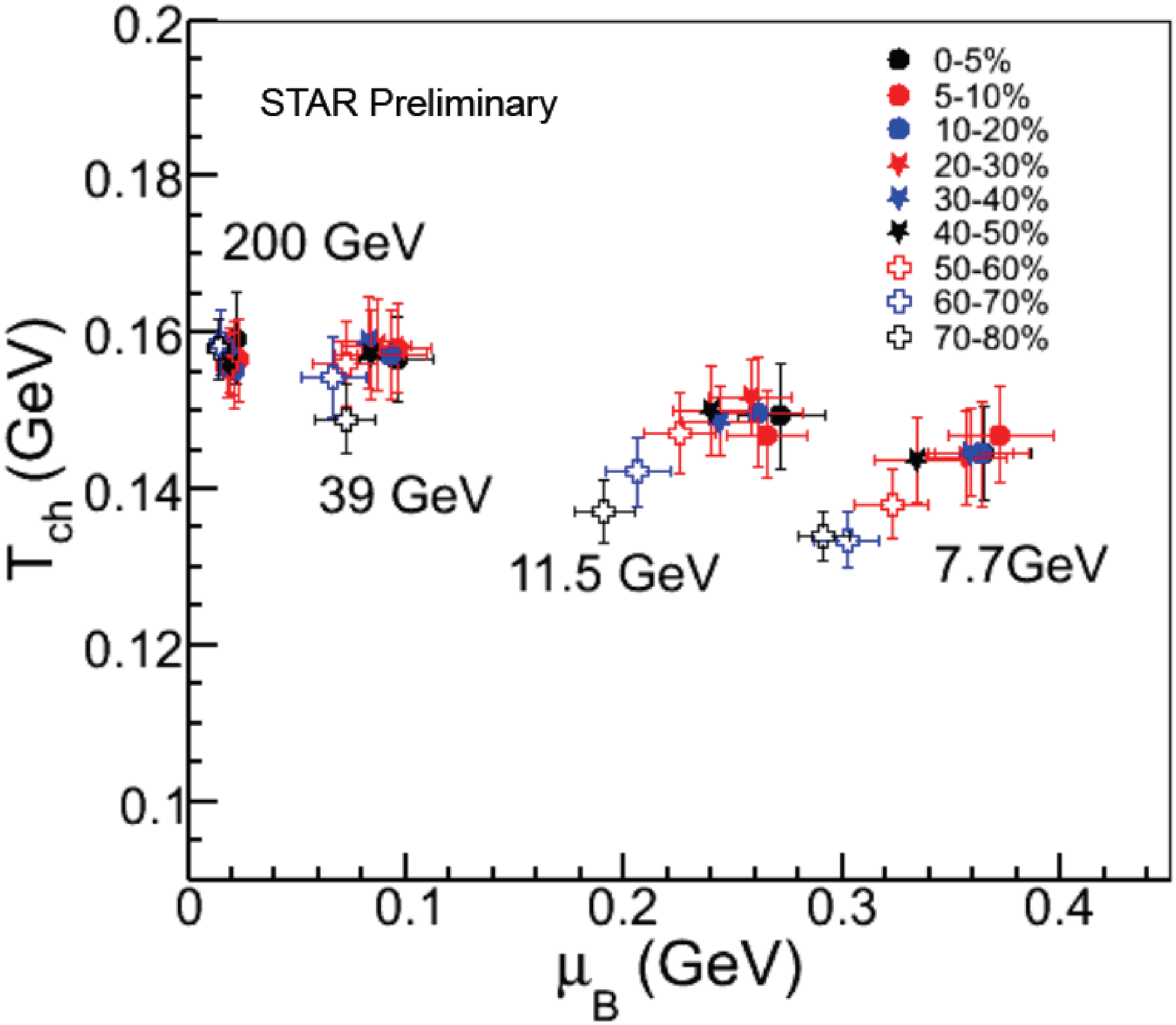} %           
\includegraphics[width=0.41\textwidth]{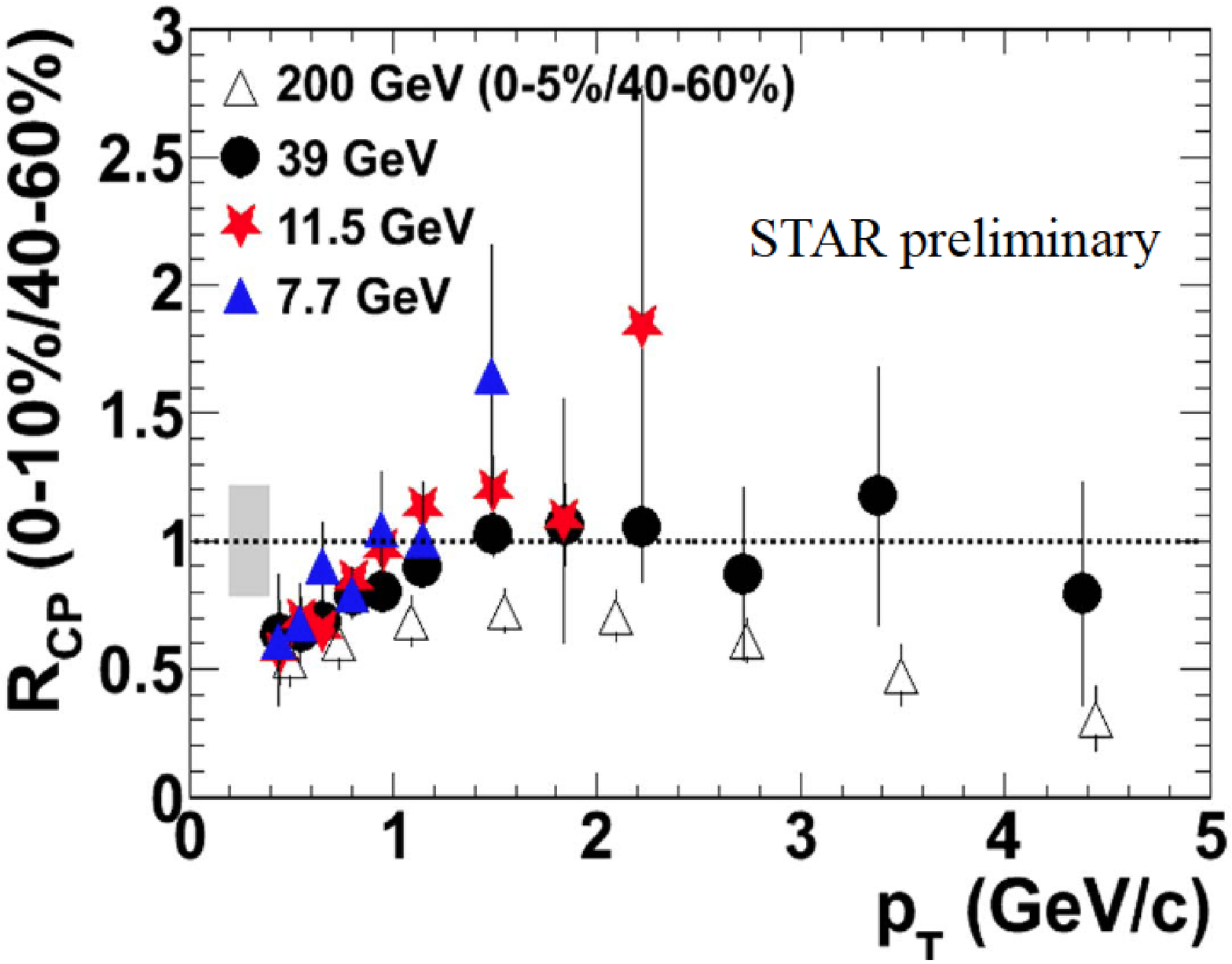} %
  \caption{(Color online) Centrality dependence of the chemical freeze-out parameters $T_{ch}$ and $\mu_{B}$ for $\sqrt{s_{NN}}$ = 7.7, 11.5, 39 and 200 GeV (left) and  $R_{\rm {CP}}$ of $\phi$ mesons for 7.7, 11.5, 39 and 200 GeV (right).}
  \label{fig_lfl}
\end{figure}

Bulk properties from heavy-ion reactions, such as the chemical and kinetic freeze-out temperatures, can be extracted from particle spectra.
Figure \ref{fig_lfl} (left) shows the chemical freeze-out parameters $T_{\rm{ch}}$ and $\mu_{B}$ for various energies and centralities, extracted from THERMUS \cite{ref_thermus} fits to particle ratio yields of kaons, pions and protons \cite{STAR_Lokesh}. For the first time a centrality dependence of these parameters is observed at the lower energies 39, 11.5 and 7.7 GeV, whereas the parameters seem to be constant at 200 GeV. This analysis will be updated soon with more particle species included. \newline
The hadronic reaction cross section of $\phi$-mesons is much lower compared to other hadrons. This makes it interesting to study the $\phi$-meson  $R_{\rm {CP}}$ for different energies above and below the phase transition line. A significant change in the R$_{\text{CP}}(p_{T})$ should then be observed between lower and higher energies. In Fig. \ref{fig_lfl} (right) the  $R_{\rm {CP}}$ of $\phi$-mesons for 7.7, 11.5, 39 and 200 GeV is shown. At 200 GeV a clear suppression at more central reactions for all transverse momenta is observed, whereas systematically higher  $R_{\rm {CP}}$ values of 39 GeV are found. The observed pattern might also be influenced by other physics like the Cronin effect. The spectra at 7.7 and 11.5 GeV follow the trend at 39 GeV within statistical error bars at transverse momenta below 1.5 and 2.2 GeV/c, respectively.

\subsection{Elliptic flow of identified particles}

\begin{figure}[htc]
  \vspace{-20pt}
\resizebox{6cm}{!}{%
  \includegraphics{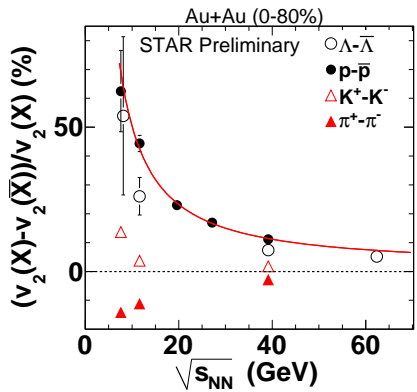}}
 \vspace{-10pt}
  \hspace{0pt}
\caption{(Color online) The difference in integrated $v_{2}$ between particles (X) and their corresponding anti-particles (${\rm \bar{X}}$) (see legend), normalized to the integrated particle $v_{2}$ as a function of $\sqrt{s_{NN}}$ for 0-80\% central Au+Au reactions.}
\label{Delta_v2_vs_sNN_CPOD}       % Give a unique label
\end{figure}

The elliptic flow $v_{2}$ is the second order harmonic of the azimuthal particle distribution relative to the reaction plane. It is generated by the pressure gradient of the overlapping nuclei, mainly during the early stage of the reaction. The number-of-constituent quark (NCQ) scaling of $v_{2}$ at top RHIC energies is interpreted as one of the signatures for the QGP phase \cite{STAR_ncq}. 
In Fig. \ref{Delta_v2_vs_sNN_CPOD} the energy dependence of the difference in the integrated $v_{2}$ between particles and their corresponding anti-particles relative to the particle $v_{2}$ is plotted \cite{STAR_Shusu}.  At higher energies we observe a difference of only  $~2\%$ between $K^{+}$ and $K^{-}$, and a difference of $-3\%$ between $\pi^{+}$ and $\pi^{-}$. An almost constant difference of $5\%$ to $10\%$ is found for baryons at 39 and 62.4 GeV. This constant behaviour changes at energies below 27 GeV. Baryons show a clear trend to larger differences with a magnitude in the relative integrated $v_{2}$ difference of up to 50\%. Pions show the opposite behaviour compared to kaons with respect to their charge. 
From the observed difference in $v_{2}$ between particles and corresponding anti-particles at lower energies, which also holds for the transverse momentum dependence, one can infer that the NCQ scaling between these particles is broken. 

\subsection{Azimuthal HBT}

\begin{figure}[htc]
  \vspace{-10pt}
\resizebox{8cm}{!}{%
  \includegraphics{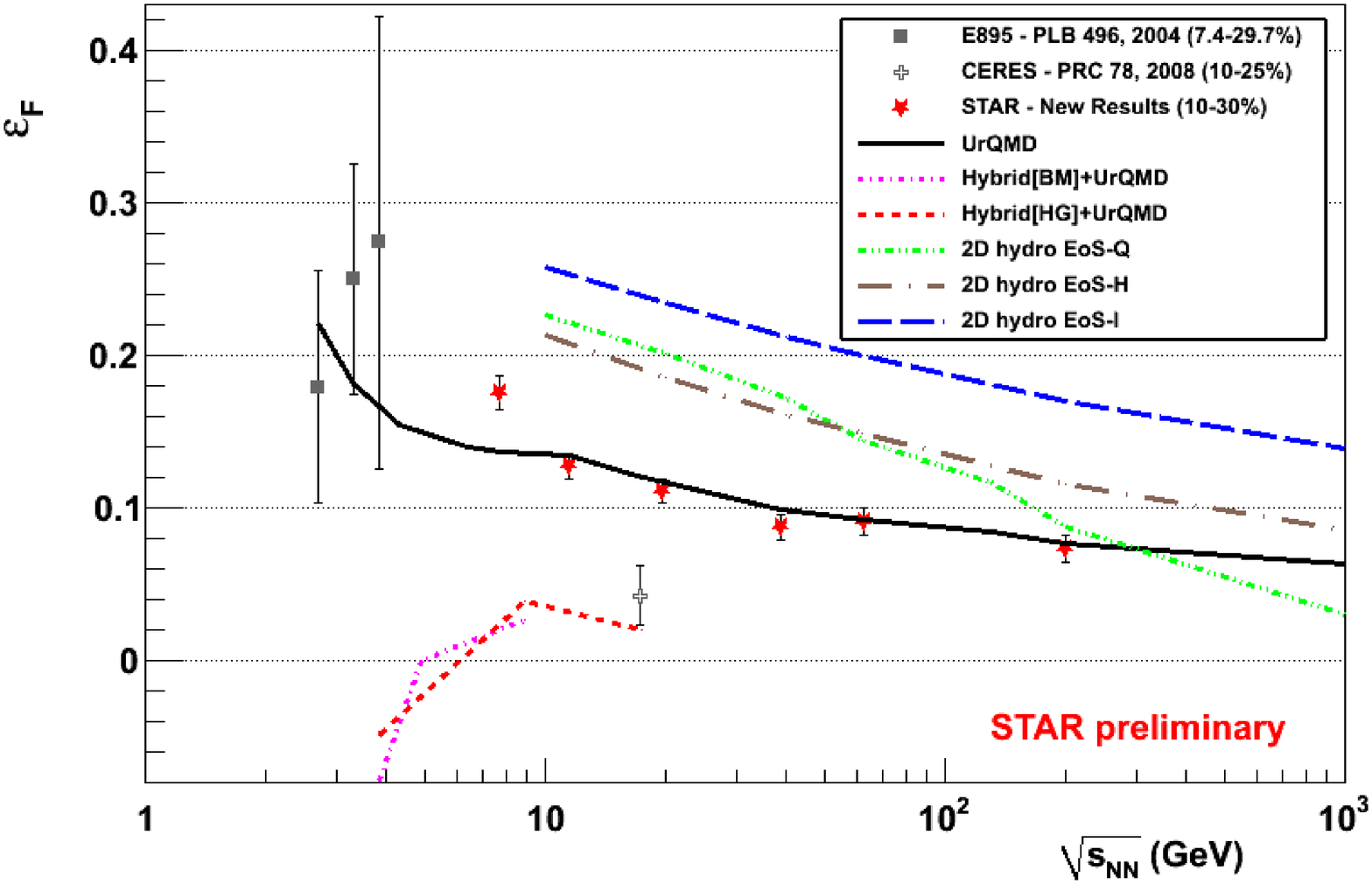}}
 \vspace{-10pt}
  \hspace{0pt}
\caption{(Color online) Freeze-out eccentricity of the charged particles as a function of $\sqrt{s_{NN}}$. The new data from STAR is shown by red star symbols. Model calculations are represented by solid and dashed lines.}
\label{AzHBT}       % Give a unique label
\end{figure}

The HBT technique can be used to determine the freeze-out eccentricity $\varepsilon_{f}=\frac{\text{R}_{y}^{2}-\text{R}_{x}^{2}}{\text{R}_{y}^{2}+\text{R}_{x}^{2}}$. This requires the HBT correlation analysis to be done as a function of the angle relative to the event plane. A non-monotonic behaviour of $\varepsilon_{f}$ as a function of energy could indicate a soft point in the equation of state. 
Figure \ref{AzHBT} shows the excitation function of the freeze-out eccentricity. Various hybrid, transport model and pure hydro calculations are overlaid \cite{STAR_Anson}. The combined E895 and STAR data show a smooth decrease of $\varepsilon_{f}$ with energy. Only the CERES data point at 17.3 GeV deviates from this behaviour. The hydro and UrQMD model predictions can describe the general trend of the decrease but only UrQMD additionally fits the magnitude of $\varepsilon_{f}$. The hybrid calculations clearly show a different trend.

\subsection{Particle ratio fluctuations}

\begin{figure}[htc]
  \hspace{20pt}
\resizebox{11cm}{!}{%
  \includegraphics{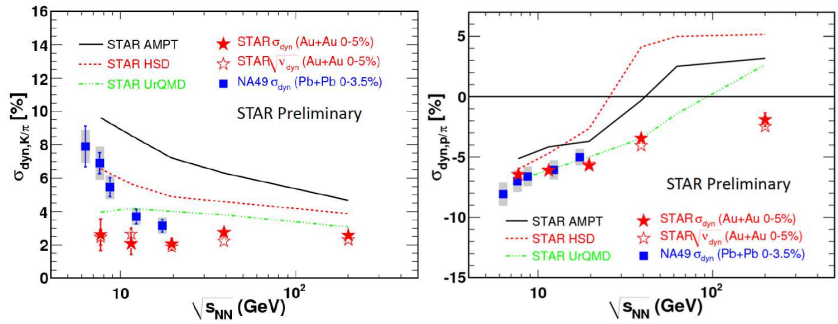}}
 \vspace{-10pt}
  \hspace{0pt}
\caption{(Color online) Energy dependence of the dynamical event-by-event particle ratio fluctuations $\sigma_{\text{dyn}}$ and $\sqrt{\nu_{\text{dyn}}}$ for $K$/$\pi$ (left figure) and $p$/$\pi$ (right figure).}
\label{Part_fluct}       % Give a unique label
\end{figure}

The existence of a QCD critical point would most probably lead to a significant increase of event-by-event fluctuations for $T-\mu_{B}$ trajectories which come close to the critical area or pass the phase transition. It was suggested to study observables which have a connection to baryon number or strangeness conservation like net-proton distributions or particle ratios like $K$/$\pi$ \cite{ref_PRF}. In the case of event-by-event particle ratios the volume may cancel out. 
The observable $\sigma_{\text{dyn}}$ \cite{ref_sigma_dyn} is used in the following discussion and is defined as:

\begin{equation}
\sigma_{\text{dyn}} = \text{sign}\left(\sigma^{2}_{\text{data}} - \sigma^{2}_{\text{mixed}}\right) \sqrt{\left| \sigma^{2}_{\text{data}} - \sigma^{2}_{\text{mixed}} \right|},
\label{form_sigmadyn}
\end{equation}

where $\sigma_{\text{data}}$ is the width of same-event distribution and $\sigma_{\text{mixed}}$ is the width of the corresponding mixed-event distribution. In addition to $\sigma_{\text{dyn}}$ the mixed event independent observable $\nu_{\text{dyn}} \approx \sigma^{2}_{\text{dyn}}$ is studied. In Fig. \ref{Part_fluct} the energy dependence of $\sigma_{\text{dyn}}$ is shown for results from the NA49 (0-3.5\%, Pb+Pb) \cite{ref_NA49} and STAR (0-5\%, Au+Au) experiments \cite{STAR_Tian}.  A fair agreement between the two data sets is reached in case of $\sigma_{\text{dyn,p}/\pi}$ whereas a significant difference is observed for $\sigma_{\text{dyn,K}/\pi}$ at lower energies. The increase of $\sigma_{\text{dyn,K}/\pi}$ with decreasing energy, observed by NA49, cannot be reproduced by the STAR results. Three model predictions are also plotted in Fig. \ref{Part_fluct}.
Higher moments of event-by-event net-proton distributions are sensitive to baryon number fluctuations. First results from an ongoing analysis do not show a discontinuity so far \cite{STAR_hm}.  

\section{Summary and outlook}

The first results from the beam energy scan program at STAR contain important new information for the search of the QCD phase transition and the critical point. Event-by-event dynamical fluctuations, particle spectra and bulk correlations like azimuthal HBT and elliptic and directed flow were studied as a function of energy and centrality. Additional beam energies will be investigated to extend the range to lower energies and to fill the gap in the $T-\mu_{B}$ plane between 11.5 and 19.6 GeV.


\begin{thebibliography}{99}
\bibitem{STAR_ncq} J.~Adams {\it et al.} (STAR Collaboration), Phys. Rev. Lett. 92, 052302 (2004)
\bibitem{STAR_BES1} B.~I.~Abelev {\it et al.} (STAR Collaboration) http://drupal.star.bnl.gov/STAR/starnotes/public/sn0493
\bibitem{STAR_BES2} M.~M.~Aggarwal {\it et al.} (STAR Collaboration) arXiv:1007.2613 (2010)
\bibitem{ref_lattice} Y.~Aoki, G.~Endrodi, Z.~Fodor, S.~D.~Katz, K.~K.~Szabo, Nature, 443 (2006) 675
\bibitem{ref_thermus} S.~Wheaton, J.~Cleymans, M.~Hauer, Computer Physics Communications, 180 (2009) 
\bibitem{STAR_Lokesh} L.~Kumar for the STAR Collaboration, these proceedings
\bibitem{STAR_Shusu}  S.~Shi for the STAR Collaboration, these proceedings
\bibitem{STAR_Anson} C.~Anson for the STAR Collaboration, arXiv:1107.1527 (2011)
\bibitem{ref_PRF} V.~Koch, A.~Majumder and J. Randrup, Phys. Rev. Lett. 95, 182301 (2005)
\bibitem{ref_sigma_dyn} S.~V.~Afanasiev {\it et al.}, Phys. Rev. Lett. 86, 1965 (2001)
\bibitem{ref_NA49} C.~Alt {\it et al.} (NA49 Collaboration), Phys. Rev. C 79, 044910 (2009)
\bibitem{STAR_Tian} J.~Tian for the STAR Collaboration, these proceedings
\bibitem{STAR_hm} X.~Luo for the STAR Collaboration, arXiv:1111.5671 (2011)
\end{thebibliography}
\end{document}